\pdfoutput=1

\documentclass[11pt]{article}

\usepackage{EMNLP2023}

\newcommand{\Ni}{(1)~}
\newcommand{\Nii}{(2)~}
\newcommand{\Niii}{(3)~}

\usepackage{xcolor}
\definecolor{Purple}{HTML}{7249AA}
\definecolor{Blue}{HTML}{276483}
\definecolor{Green}{HTML}{206B2D}%
\definecolor{Gray}{HTML}{5E5E5E}
\definecolor{Background}{HTML}{FDF6E3}
\definecolor{LightGray}{gray}{0.9}

\usepackage{graphicx}
\usepackage{times}
\usepackage{latexsym}
\usepackage[T1]{fontenc}
\usepackage[utf8]{inputenc}
\usepackage{microtype}
\usepackage{xspace}
\usepackage{enumitem}
\usepackage{booktabs}
\usepackage{amsmath}
\usepackage{pifont}
\usepackage{multirow}
\usepackage{tablefootnote}
\usepackage{graphicx}
\usepackage{listings}
\usepackage[hang,flushmargin]{footmisc}
\lstdefinestyle{custom-style}{
    language=Python,
    backgroundcolor=\color{Background},   
    commentstyle=\color{Green},
    keywordstyle=\bfseries\color{Blue},
    numberstyle=\tiny\color{Gray},
    stringstyle=\color{Purple},
    basicstyle=\footnotesize\ttfamily\color{black},
    frame=lines,
    breakatwhitespace=false,         
    breaklines=true,                 
    captionpos=b,                    
    keepspaces=true,                 
    numbersep=5pt,                  
    showspaces=false,                
    showstringspaces=false,
    showtabs=false,                  
    tabsize=1
}

\usepackage{inconsolata}

\newcommand{\Spacerini}{~Spacerini\xspace}
\newcommand{\github}{\hspace{0em}\raisebox{-0.2ex}{\includegraphics[width=1em]{figures/github-logo.pdf}}\hspace{0.05em}}
\renewcommand{\github}{gh}
\newcommand{\huggingface}{\hspace{0em}\raisebox{-0.2ex}{\includegraphics[width=1em]{figures/hf-logo.pdf}}\hspace{0.05em}}
\renewcommand{\huggingface}{hf}
\newcommand{\hflink}[2]{\texttt{\mbox{\href{#1}{\huggingface/#2}}}}
\newcommand{\ghlink}[2]{\texttt{\mbox{\href{#1}{\github/#2}}}}

\newcommand{\ignore}[1]{}

\begin{document}

\title{Spacerini: Plug-and-play Search Engines with Pyserini and Hugging Face}

\author{
Christopher Akiki\textsuperscript{1,2} 
\\\And Odunayo Ogundepo\textsuperscript{3}
\\\And Aleksandra Piktus\textsuperscript{4,5}
\\\AND Xinyu Zhang\textsuperscript{3}
\\\And Akintunde Oladipo\textsuperscript{3}
\\\And Jimmy Lin\textsuperscript{3}
\\\And Martin Potthast\textsuperscript{1,2}
\\\AND\normalfont
\textsuperscript{1} Leipzig University $\quad$
\textsuperscript{2} ScaDS.AI $\quad$
\textsuperscript{3} University of Waterloo$\quad$ \\
\textsuperscript{4} Hugging Face$\quad$
\textsuperscript{5} Sapienza University $\quad$
}
  
\maketitle

\begin{abstract}
We present \Spacerini, a tool that integrates the Pyserini toolkit for reproducible information retrieval research with Hugging Face to enable the seamless construction and deployment of interactive search engines. \Spacerini makes state-of-the-art sparse and dense retrieval models more accessible to non-IR practitioners while minimizing deployment effort. This is useful for NLP~researchers who want to better understand and validate their research by performing qualitative analyses of training corpora, for IR~researchers who want to demonstrate new retrieval models integrated into the growing Pyserini ecosystem, and for third parties reproducing the work of other researchers. \Spacerini is open-source%
\footnote{\url{https://github.com/castorini/hf-spacerini}}
and includes utilities for loading, preprocessing, indexing, and deploying search engines locally and remotely. We demonstrate a portfolio of 13~search engines created with \Spacerini for different use cases.%
\footnote{\url{https://hf.co/spacerini}}
\end{abstract}
       
\section{Introduction}
\label{sec:intro}

\enlargethispage{\baselineskip}
The commoditization of data has transformed data-driven computer science in general \citep{hey2009-fourth-paradigm} and machine learning (ML) and natural language processing (NLP) in particular \citep{mitchell-measuring-data-22}. The race to train ever-larger language models depends so much on having access to immense amounts of text \citep{chinchilla-hoffmann-22} that the datasets have become, as \citet{bender-gebru-stochastic-21} claims, both technically \citep{bender-gebru-stochastic-21} and methodologically \citep{jo-archives-20} ``too big to document''. In practice, this often leads to an approach of training first and asking questions later  \citep{akiki-bigscience-22}, which again is an example of a convenience experiment \citep{krohs-convenience-experimentation}, a research approach that depends on the availability of a resource and the ease of a method rather than its suitability to the problem at hand. In this sense, \citet{beaulieu-leonelli-2021-data} believe it is important to distinguish between the availability of data and its appropriateness, especially in light of the misconception that web data represents all human experience and is immune to the ever-widening \emph{digital divide} \citep{leonelli-science-big-data}. They suggest that the divide not only exists, but limits the representativeness of the web, which in turn reinforces the biases of the artifacts that use it \citep{bender-gebru-stochastic-21}.

\begin{figure*}[t]
\centering
\fbox{
\includegraphics[width=0.97\textwidth]{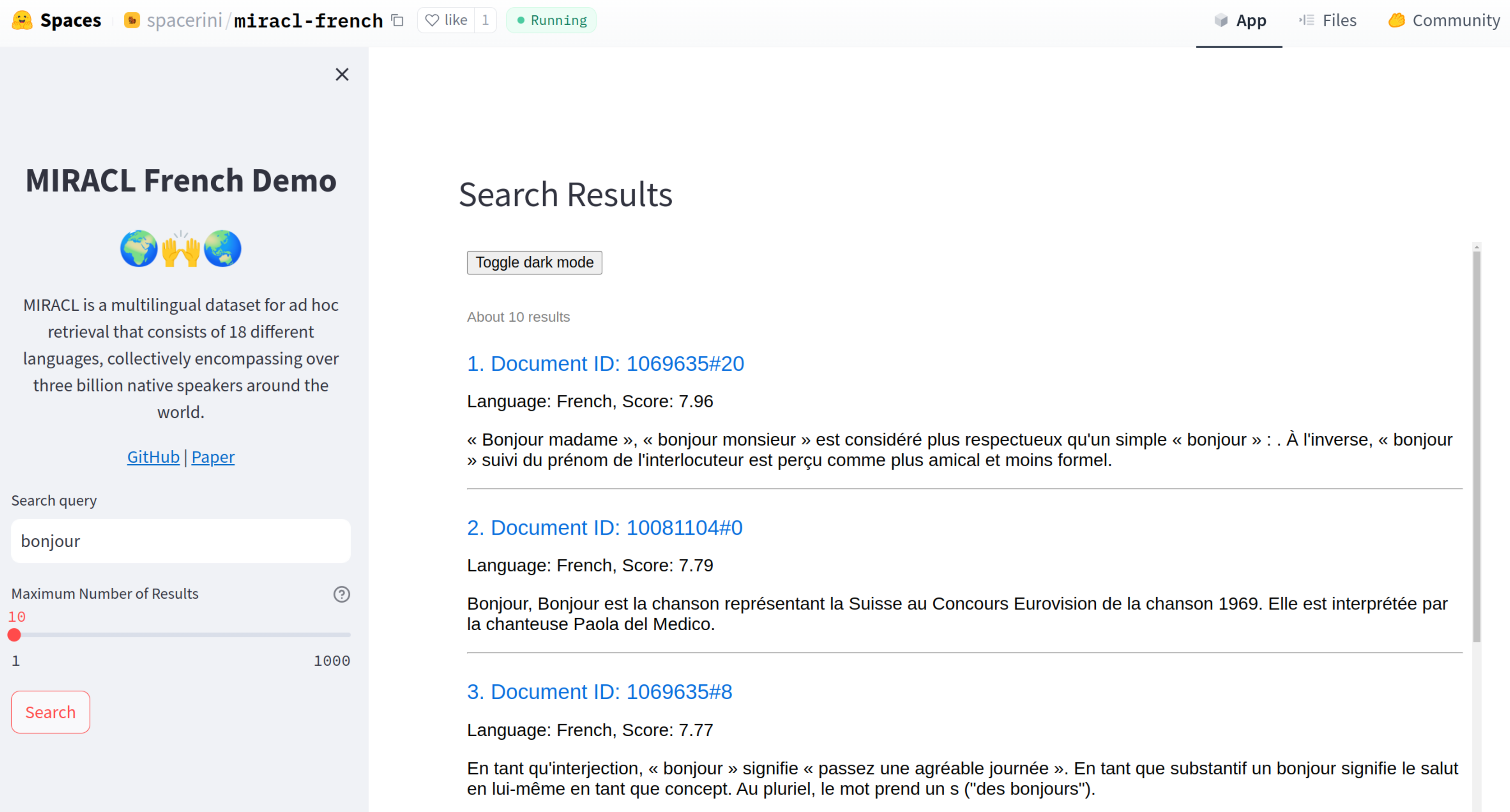}
}
\caption{Example of one of the many search apps (\hflink{https://huggingface.co/spaces/spacerini/miracl-french}{spacerini/miracl-french}) deployed as a Hugging Face Space. The Lucene BM25 index is hosted in the same repository as the frontend using \texttt{git LFS} and the frontend is based on a \texttt{template} which was automatically generated from one of the many \Spacerini \texttt{cookiecutter} templates. The loading, preprocessing, indexing, and app deployment were made using an end-to-end workflow similar to the one showcased in Section~\ref{sec:package}.}
\label{fig:search_screenshot}
\end{figure*}

Being unable to easily audit large datasets incentivizes researchers to release models trained on data they do not truly understand \citep{mitchell-measuring-data-22} leading to model behaviors that are hard to study, predict or trace \citep{mitchell-measuring-data-22,hooker-archaelogy-22,akyurek-etal-2022-towards}. This is especially problematic in light of the potential real-world harms that ensue \citep{llm-risks-harms,hutchinson-etal-2020-social,twitter-abusive-behavior-crowdsourcing,bias-online-fiction-writing}. Being able to properly understand the limitations of our datasets and qualitatively explore them in an ad-hoc fashion is a necessary first step toward understanding the behavior, harmful biases and failure modes of the artifacts that build upon them. Understanding the training data is therefore a critical step in the process of releasing and auditing large language models \citep{auditing-llms}.

It is from this vantage point that we initially developed \Spacerini as an open-source tool for the quick indexing and deployment of shareable search engines, but have also since come to realize its potential in being useful for an even wider audience interested in making their text artifacts searchable. Indeed this includes IR students, Digital Humanists, Shared Tasks organizers, and digital investigative journalists, all of whom seek to quickly deploy ad hoc search engines for their research. We cover these use cases in more detail in Section~\ref{sec:use-cases}. 

\Spacerini helps streamline the process of auditing large datasets by allowing users to effortlessly index their text collections and deploy them as interactive search applications that can be easily edited in the browser and shared with all stakeholders. It achieves that by ``standing on the shoulders'' of battle-tested open-source libraries from the Castorini \citep{Lin_etal_SIGIR2021_Pyserini}, Hugging Face \citep{abid-2019-gradio,huggingface-datasets-lhoest-21} and Python ecosystems. It also enhances the interoperability between them to enable quick indexing and free deployment of search interfaces in an easy-to-use package that makes it possible to reduce the overhead typically involved in operationalizing data governance frameworks, and allows stakeholders to focus on data analysis rather than data engineering. This is achieved through the modularity of its design that enables data loading~(Section~\ref{sec:package:loading}), preprocessing~(Section~\ref{sec:package:preprocessing}), dense and sparse indexing~(Sections~\ref{sec:package:indexing}), as well as the creation~(Section~\ref{sec:package:loading}), and free hosting of graphical search interfaces~(Section~\ref{sec:package:spaces}) for text datasets. 

\section{Background and Related Work}

Large scale, predominantly web-mined text datasets have been proliferating in NLP recently, giving rise to publications \cite{laurencon2022the, gao-pile-21,suarez-processing-huge-corpora,t5-raffel} which often contain interesting analyses of the specific datasets being presented, however, usually lack any comparison to existing resources beyond basic metrics such as sizes of the datasets or languages they contain. 

As discussed in Section~\ref{sec:intro}, in the face of an increased scrutiny of the models trained on datasets in question, the topic of data understanding and governance has been gaining more traction, being accepted as an important part of research. Efforts such as those of \citet{mitchell-measuring-data-22} contribute frameworks for more standardised and reproducible metrics and measurements of datasets, and we position ourselves as a complementary continuation of their work, focusing on a more curatorial and qualitative assessment that might not readily fit under the umbrella of ``measurements''. We therefore aim to fill the gap in the evaluation landscape by facilitating qualitative, rather than quantitative analysis of large scale data collections.

Similarly to the authors of Gradio \cite{abid-2019-gradio}, a Python package for fast development of Machine Learning demos, we believe that the accessibility of data and model analysis tools is crucial to building both the understanding of and the trust in the underlying resources.
The potential of relevance-based interfaces to massive textual corpora, the creation of which can be facilitated by leveraging  toolkits such as Pyserini \cite{Lin_etal_SIGIR2021_Pyserini},
 has previously been tapped into by the researchers at the Allen Institute of AI who propose a C4 \cite{t5-raffel} search engine\footnote{\url{https://c4-search.apps.allenai.org/}}. 
 Similar interfaces have also been found useful in more specialised domains, e.g. in COVID-related datasets \citep{zhang2020covidex}, news quotes \cite{Vukovi__2022}, or medical literature \citep{cancer-nlp-no-code}. However, while these solutions are undeniably useful, they remain very contextual: dataset-specific, and project-specific. We believe \Spacerini to be the first generalizable tool which proposes an end-to-end pipeline automating the route from raw text to qualitative analysis.
\section{Spacerini}
\label{sec:package}
\Spacerini is a modular framework that integrates Pyserini with the Hugging Face ecosystem to streamline the process of going from any Hugging Face text dataset (or indeed any text dataset)---either local or hosted on the Hugging Face Hub---to a search interface driven by a Pyserini index that can be deployed for free on the Hugging Face Hub. In what follows, we deconstruct an example script\footnote{\url{https://github.com/castorini/hf-spacerini/blob/main/examples/scripts/gradio-demo.py}} to showcase the different features enabled by \Spacerini. When run end-to-end, the script pulls a dataset from Hugging Face, pre-processes it, indexes it, creates a gradio-based search interface and deploys that as a Hugging Face Spaces demo. This is only meant as a feature-complete demo, and we don't expect most people to want to integrate every step into their workflows, but rather to cherry-pick and decide what best to use depending on context.

\subsection{Loading Data}
\label{sec:package:loading}
All our workflows are backed by the Hugging Face \texttt{datasets} library \citep{huggingface-datasets-lhoest-21}, itself based on the extremely efficient Apache Arrow format. \texttt{Datasets} is a mature library which provides a standardized interface to any tabular dataset, in particular, to tens of thousands of community datasets hosted on the Hugging Face Hub\footnote{\url{https://hf.co/datasets}}. The \texttt{datasets} library gives fine-grained control over the lifecycle of tabular datasets, which we choose to abstract away through a set of opinionated data loading functions that cover the use cases we deem relevant to information retrieval. We also add new functionality, such as the ability to load any document dataset from the \texttt{ir\_datasets} library using for example the following one-liner to load MS~MARCO \citep{msmarco} as a Hugging Face \texttt{datasets.Dataset} object using a function from the \texttt{data} subpackage:
\begin{lstlisting}
from spacerini.data import load_ir_dataset

hf_dset = load_ir_dataset("msmarco-passage")
\end{lstlisting}
We include wrappers to load database tables, pandas DataFrames \citep{pandas-software,pandas-paper}, and text datasets on disk, as well as the ability to load any dataset either in memory-mapped mode or in streaming mode: the former makes it possible to handle larger-than-memory datasets, and the latter larger-than-disk datasets that can be streamed from a remote location such as the Hugging Face Hub.

\subsection{Pre-processing}
\label{sec:package:preprocessing}
\Spacerini also provices a \texttt{preprocess} subpackage which offers a range of customizable pre-processing options for preparing datasets.
This module includes a sharding utility that enables the partitioning of large datasets into smaller, more manageable chunks for efficient parallel processing.
\begin{lstlisting}
from spacerini.preprocesss import shard_dataset
shard_dataset(
    hf_dataset=hf_dset,               
    shard_size="1GB",
    column_to_index="text", 
    shards_paths="msmarco-shards",
)
\end{lstlisting}
\subsection{Indexing}
\label{sec:package:indexing}
\Spacerini's \texttt{index} subpackage leverages Pyserini to provide very efficient Lucene indexing and allow users to easily and quickly index large datasets, either sharded in the pre-processing step, or any text format accepted by Pyserini, and streaming text datasets, such as those returned by \Spacerini's \texttt{data} subpackage. This subpackage also exposes several tokenization options using existing language-specific analyzers\footnote{\url{https://lucene.apache.org/core/9_5_0/analysis/common/}} as well as Hugging Face subword tokenizers \citep{moi-hf-tokenizers}. 

\begin{lstlisting}
from spacerini.index import index_json_shards

index_json_shards(
    shards_path="msmarco-shards",
    index_path="app/index",
)
\end{lstlisting}

We also provide wrappers to Pyserini's dense and hybrid retrieval functionality through the \texttt{spacerini.index.encode} subpackage.

\subsection{Template-based Search Interfaces}
\label{sec:package:cookiecutter}
Having indexed a collection, one can easily spin up a frontend using the \texttt{frontend} subpackage and one of many provided templates\footnote{\mbox{\url{https://github.com/castorini/hf-spacerini/tree/main/templates}}\hspace*{-3pt}}. These are built using \texttt{cookiecutter},\footnote{\url{https://github.com/cookiecutter/cookiecutter}} a Python templating library for software projects. We provide a few batteries-included frontend templates based both on the Gradio\footnote{\url{https://gradio.app/}} and Streamlit \footnote{\url{https://streamlit.io/}} demo app frameworks, both of which are natively supported by Hugging Face \texttt{Spaces}. Figure~\ref{fig:search_screenshot} showcases a search engine built using one of our Streamlit templates.
\begin{lstlisting}
from spacerini.frontend import create_app

cookiecutter_vars = {
    "dset_text_field": "text",
    "metadata_field": "docid",
    "space_title": "MS MARCO Search",
    "local_app": "app"
}
create_app(
    template="gradio-vanilla",
    extra_context_dict=cookiecutter_vars,
    output_dir=".",
)
\end{lstlisting}
\subsection{Deployment to Hugging Face Spaces}
\label{sec:package:spaces}
The local apps developed in the previous subsection can then be pushed to Hugging Face Spaces and hosted there for free. One can then further customize the running app from the browser, for example to add functionality not provided by the chosen template. The goal of the templates is to provide a useful starting point in the form of a running app that users can further customize with interface features useful for their own workflows.

\begin{lstlisting}
from spacerini.frontend import create_space_from_local

create_space_from_local(
    space_slug="msmarco-passage-search,
    organization="spacerini,
    space_sdk="gradio",
    local_dir=LOCAL_APP,
    delete_after_push=True,
)
\end{lstlisting}

\subsection{Sharing Indexes as Hugging Face Datasets}
\label{sec:index-sharing}
Orthogonal to the workflow presented so far, is the ability to upload Lucene indexes to the Hugging Face Hub using shareable dataset repositories and enabling reproducible retrieval experiments.
\begin{lstlisting}

from spacerini.index import push_index_to_hub

push_index_to_hub(
    dataset_slug="lucene-english-analyzer-msmarco",
    index_path="index",
)
\end{lstlisting}

Any hosted index can then just as easily be downloaded for local use:
\begin{lstlisting}

from spacerini.index import load_index_from_hub

index_path = load_index_from_hub("lucene-fr-analyzer-")
\end{lstlisting}

\subsection{Search and Pagination}
Search features are provided by the \texttt{search} subpackage and leverage the memory-mapping feature of Arrow tables to load the entire table of results---no matter how big---only materializing the specific shard that corresponds to the requested result page.
\begin{lstlisting}
from spacerini.search import result_indices, result_page 

ix = result_indices(
    "Lorem Ipsum",
    num_results=1_000,
    INDEX_PATH,
)

last_results_page = result_page(
    hf_dset,
    ix,
    page=-1,
    results_per_page=20,
).to_pandas()    
\end{lstlisting}

\section{Use Cases and Demonstrations}
\label{sec:use-cases}

We envision \Spacerini to be useful primarily to NLP researchers, students, shared task organizers, data scientists, and data annotators, as well as tech-adjacent and -proficient professionals and laypeople. In what follows, we overview a series of 13~use-cases that we implemented and how they might benefit their respective targeted audience. An overview of all 13~search engines can be found at \url{https://huggingface.co/spacerini}; in following inline links to the selected engines, this part of their URL prefix is shortened to `{\tt \href{https://huggingface.co/spacerini}{hf}}' for brevity.

\subsection{NLP researchers}
\Spacerini is designed to enable qualitative analysis of large-scale textual corpora without the need for extensive engineering work. It can be used in dataset auditing campaigns, such as those carried out by \citet{kreutzer-etal-2022-quality} or in data annotation efforts. Also relevant here are generative text models whose outputs can be better understood by better understanding the datasets they were trained on. We refer the reader to Section~5 of \citet{piktus-etal-2023-roots} for a detailed exposé of potential use scenarios which include: PII detection, problematic content detection, social representation, benchmark and language contamination detection, as well as plagiarism and memorization detection. An example demo for this context is the index of the XSUM \citep{xsum-emnlp} dataset which is indexed and can be explored with the \hflink{https://huggingface.co/spaces/spacerini/xsum-search}{xsum-search} demo. 

\subsection{IR researchers}
Given its tight integration with Pyserini, \Spacerini can also be leveraged by IR researchers to experiment with modifications of their retrieval pipelines in user studies or to deploy demos of their working prototypes. Reproducibility for IR experiments is further enhanced thanks to the index sharing abilities showcased in Section~\ref{sec:index-sharing}. As a practical example, \Spacerini was used in the context of the BigCode project \citep{starcoder} to quickly experiment with multiple n-gram tokenization schemes for BM25-based code retrieval; this corresponds to the \hflink{https://huggingface.co/spaces/spacerinicode-search}{spacerini/code-search} demo.

\subsection{Linguists}
Corpus linguistics relies on qualitative analyses of text corpora to understand language and its many varieties by studying the way it is used \citep{mcenery_hardie_2011,piktus-etal-2023-roots}. Some of these empirical analyses can be facilitated by the usage of an inverted index, coupled with the correct querying patterns and frontend elements, both of which are easy to achieve using \Spacerini. 

\subsection{Digital Humanists}
\Spacerini can also be leveraged by Digital and Computational Humanists, Archivists and Librarians looking to index their collections. Indeed, GLAM~(Galleries, libraries, archives, and museums) collections are increasingly being made available as datasets. Furthermore, there is a growing interest in the digital humanities in training and using languages models, as demonstrated by the success of projects such as the \emph{AI for Humanists Project}.\footnote{\url{https://www.bertforhumanists.org/}} In this context, indexing data relevant to these efforts is a difficult task; often project-based and contingent upon precarious funding arrangements. Having a project-agnostic tool like \Spacerini could prove valuable to this community and a useful addition to toolkits such as the GLAM Workbench \citep{glam-workbench}.

\subsection{IR Students}
Given its low barrier of entry, \Spacerini can be a good tool for IR courses, where students could be tasked with developing search engines, by providing an easy-to-deploy frontend interface for their developed retrieval systems which does not even have to be deployed within the same application, as demonstrated by the \hflink{https://huggingface.co/spaces/spacerini/chat-noir}{chat-noir}, a frontend wrapper for ChatNoir \citep{bevendorff:2018}.

\subsection{Shared task organizers}
\Spacerini can also be leveraged by organizers of shared tasks such as MIRACL \cite{zhang-miracl-22} and Touch\'{e} \citep{touche-22}, who want to help participants explore the datasets without forcing them to download large volumes of data or giving participants full access to the data: it is indeed possible to host the index privately on the Hugging Face Hub and only expose access to it through a search interface. \Spacerini can also be used as a platform for participants to deploy working prototypes of their submissions with a unified interface provided by the organizer as a \texttt{cookiecutter} template. Example demos for this use case include  \hflink{https://huggingface.co/spaces/spacerini/miracl-bengali}{miracl-bengali}, \hflink{https://huggingface.co/spaces/spacerini/miracl-arabic}{miracl-arabic}, and \hflink{https://huggingface.co/spaces/spacerini/miracl-swahili}{miracl-swahili}.

\subsection{Tech journalists}
\Spacerini can help data journalists and digital investigative journalists index, explore, and understand open data, in a similar vein to the functionality provided by the Aleph suite.\footnote{\url{https://docs.alephdata.org/}} Providing technical tools to data journalists is a crucial in uncovering matters of public interest, as was evident by role played by the collaborative use of the Neo4j graph database in unraveling corrupt networks surrounding tax havens \citep{neo4j-panama}.

\subsection{Additional usage patterns}
Finally, three features of Hugging Face Spaces make them especially attractive for users: \Ni~they can leverage private datasets, meaning that one can provide search access to a dataset without sharing the underlying data,\Nii~they can be seamlessly embedded into HTML, specifically Gradio-based Spaces which can be embedded as \emph{Web Components}\footnote{\url{https://developer.mozilla.org/en-US/docs/Web/Web_Components}} so that users can easily integrate a \Spacerini-based search feature into their own websites\footnote{e.g.: \url{https://cakiki.github.io/search-engine/}}, and \Niii~Gradio-based Spaces expose a FastAPI\footnote{\url{https://github.com/tiangolo/fastapi}} endpoint that can be queried to access the functionality of the space, making deployed search engines accessible through HTTP~calls.
\section{Limitations and Future Plans}

The main limitation of the off-the-shelf variant of \Spacerini is the disk space limit imposed by Hugging Face Spaces, which is currently set to 50~GB for the free tier.\footnote{\url{https://huggingface.co/docs/hub/spaces-overview\#hardware-resources}}
While not enough to accommodate entire corpora such as ROOTS or The Pile, such datasets are typically amalgamations of constituent datasets which can each be studied independently. This limit has no bearing on \Spacerini search apps deployed locally. Should users still want to get more disk space for their Spaces-hosted indexes, they can either pay for an upgrade to a more appropriate tier or see whether they qualify for a free hardware upgrade through the community grants offered by Hugging Face, in the \texttt{Settings} pane of the relevant space.

Planned improvements include automating the creation of dataset cards (or rather ``index cards'') when pushing an index to the Hugging Face Hub, better documentation, as well as more fine-grained tokenization support.

Please also note that \Spacerini is currently in active development and that the stability of its current API and subpackages isn't guaranteed not to involve breaking changes as we converge toward the first stable version release. We also look forward to community contributions both to the codebase and to the frontend templates, as well as in the form of actual use cases of the library that culminate in the deployment of search apps.

\section{Conclusion}

We presented \Spacerini, a modular framework that enables the quick and free deployment and serving of template-based search indexes as interactive applications for ad-hoc exploration of text datasets. The need for such a tool is especially pressing as large language models have come to consume inordinate amounts of text data, reinforcing the need for a qualitative exploration and understanding of datasets to assess them in a way that is impenetrable to quantitative analyses alone.

\Spacerini leverages features from both the \texttt{Pyserini} toolkit and the Hugging Face ecosystem to facilitate the creation and hosting of user-friendly search systems for text datasets. Users can easily index their collections and deploy them as ad-hoc search interfaces, making the retrieval of relevant data points a quick and efficient process. The user-friendly interface enables non-technical users to effectively search massive datasets, making \Spacerini a valuable tool for anyone looking to audit their text collections qualitatively. The framework is open-source and available on GitHub under
\ghlink{https://github.com/castorini/hf-spacerini}{castorini/hf-spacerini} and demo search apps can be found under \hflink{https://huggingface.co/spacerini}{spacerini}

The key advantage of \Spacerini is its ability to simplify the search process, allowing researchers to conduct quick and efficient audits, while abstracting away all the minutiae of indexing data or hosting services. We believe that this provides an opportunity for collaboration and transparency in IR and NLP research. With the creation and sharing of search indexes publicly, practitioners, researchers and the general public can work together to pinpoint problematic content, find duplicates, and identify biases in datasets. 

Finally, we emphasize that \Spacerini is a first step in the direction of systematic dataset auditing, and more work is still needed to create standardized structures that leverage tools such as ours to properly document the different axes of interest.

\section*{Acknowledgments}
We are grateful to Lukas Gienapp, Daniel van Strien, Yuvraj Sharma, Omar Sanseviero, Julien Chaumond, Lucain Pouget, Abubakar Abid, and Leandro von Werra for their tireless support and invaluable advice throughout this project.

\begin{raggedright}
\bibliographystyle{acl_natbib}
\bibliography{emnlp23-demo-hf-spacerini-lit}

\begin{thebibliography}{40}
\expandafter\ifx\csname natexlab\endcsname\relax\def\natexlab#1{#1}\fi

\bibitem[{Abid et~al.(2019)Abid, Abdalla, Abid, Khan, Alfozan, and
  Zou}]{abid-2019-gradio}
Abubakar Abid, Ali Abdalla, Ali Abid, Dawood Khan, Abdulrahman Alfozan, and
  James Zou. 2019.
\newblock Gradio: Hassle-free sharing and testing of ml models in the wild.
\newblock \emph{arXiv preprint arXiv:1906.02569}.

\bibitem[{Akiki et~al.(2022)Akiki, Pistilli, Mieskes, Gall{\'{e}}, Wolf, Ilic,
  and Jernite}]{akiki-bigscience-22}
Christopher Akiki, Giada Pistilli, Margot Mieskes, Matthias Gall{\'{e}}, Thomas
  Wolf, Suzana Ilic, and Yacine Jernite. 2022.
\newblock \href {https://doi.org/10.48550/arXiv.2212.04960} {Bigscience: {A}
  case study in the social construction of a multilingual large language
  model}.
\newblock \emph{CoRR}, abs/2212.04960.

\bibitem[{Akyurek et~al.(2022)Akyurek, Bolukbasi, Liu, Xiong, Tenney, Andreas,
  and Guu}]{akyurek-etal-2022-towards}
Ekin Akyurek, Tolga Bolukbasi, Frederick Liu, Binbin Xiong, Ian Tenney, Jacob
  Andreas, and Kelvin Guu. 2022.
\newblock \href {https://aclanthology.org/2022.findings-emnlp.180} {Towards
  tracing knowledge in language models back to the training data}.
\newblock In \emph{Findings of the Association for Computational Linguistics:
  EMNLP 2022}, pages 2429--2446, Abu Dhabi, United Arab Emirates. Association
  for Computational Linguistics.

\bibitem[{Beaulieu and Leonelli(2021)}]{beaulieu-leonelli-2021-data}
Anne Beaulieu and Sabina Leonelli. 2021.
\newblock \emph{Data and Society: A Critical Introduction}.
\newblock Sage.

\bibitem[{Bender et~al.(2021)Bender, Gebru, McMillan-Major, and
  Shmitchell}]{bender-gebru-stochastic-21}
Emily~M. Bender, Timnit Gebru, Angelina McMillan-Major, and Shmargaret
  Shmitchell. 2021.
\newblock \href {https://doi.org/10.1145/3442188.3445922} {On the dangers of
  stochastic parrots: Can language models be too big?}
\newblock In \emph{Proceedings of the 2021 ACM Conference on Fairness,
  Accountability, and Transparency}, FAccT '21, page 610–623, New York, NY,
  USA. Association for Computing Machinery.

\bibitem[{Bevendorff et~al.(2018)Bevendorff, Stein, Hagen, and
  Potthast}]{bevendorff:2018}
Janek Bevendorff, Benno Stein, Matthias Hagen, and Martin Potthast. 2018.
\newblock {Elastic ChatNoir: Search Engine for the ClueWeb and the Common
  Crawl}.
\newblock In \emph{Advances in Information Retrieval. 40th European Conference
  on IR Research (ECIR 2018)}, Lecture Notes in Computer Science, Berlin
  Heidelberg New York. Springer.

\bibitem[{Bondarenko et~al.(2022)Bondarenko, Fr{\"{o}}be, Kiesel, Syed, Gurcke,
  Beloucif, Panchenko, Biemann, Stein, Wachsmuth, Potthast, and
  Hagen}]{touche-22}
Alexander Bondarenko, Maik Fr{\"{o}}be, Johannes Kiesel, Shahbaz Syed, Timon
  Gurcke, Meriem Beloucif, Alexander Panchenko, Chris Biemann, Benno Stein,
  Henning Wachsmuth, Martin Potthast, and Matthias Hagen. 2022.
\newblock \href {https://doi.org/10.1007/978-3-031-13643-6\_21} {Overview of
  touch{\'{e}} 2022: Argument retrieval}.
\newblock In \emph{Experimental {IR} Meets Multilinguality, Multimodality, and
  Interaction - 13th International Conference of the {CLEF} Association, {CLEF}
  2022, Bologna, Italy, September 5-8, 2022, Proceedings}, volume 13390 of
  \emph{Lecture Notes in Computer Science}, pages 311--336. Springer.

\bibitem[{D{\'i}az-Struck and Cabra(2018)}]{neo4j-panama}
Emilia D{\'i}az-Struck and Mar Cabra. 2018.
\newblock \href {https://doi.org/10.1007/978-3-319-97283-1_6} {\emph{Uncovering
  International Stories with Data and Collaboration}}, pages 55--65. Springer
  International Publishing, Cham.

\bibitem[{Fast et~al.(2016)Fast, Vachovsky, and
  Bernstein}]{bias-online-fiction-writing}
Ethan Fast, Tina Vachovsky, and Michael~S. Bernstein. 2016.
\newblock \href {http://arxiv.org/abs/1603.08832} {Shirtless and dangerous:
  Quantifying linguistic signals of gender bias in an online fiction writing
  community}.
\newblock \emph{CoRR}, abs/1603.08832.

\bibitem[{Founta et~al.(2018)Founta, Djouvas, Chatzakou, Leontiadis, Blackburn,
  Stringhini, Vakali, Sirivianos, and
  Kourtellis}]{twitter-abusive-behavior-crowdsourcing}
Antigoni{-}Maria Founta, Constantinos Djouvas, Despoina Chatzakou, Ilias
  Leontiadis, Jeremy Blackburn, Gianluca Stringhini, Athena Vakali, Michael
  Sirivianos, and Nicolas Kourtellis. 2018.
\newblock \href {http://arxiv.org/abs/1802.00393} {Large scale crowdsourcing
  and characterization of twitter abusive behavior}.
\newblock \emph{CoRR}, abs/1802.00393.

\bibitem[{Gao et~al.(2021)Gao, Biderman, Black, Golding, Hoppe, Foster, Phang,
  He, Thite, Nabeshima, Presser, and Leahy}]{gao-pile-21}
Leo Gao, Stella Biderman, Sid Black, Laurence Golding, Travis Hoppe, Charles
  Foster, Jason Phang, Horace He, Anish Thite, Noa Nabeshima, Shawn Presser,
  and Connor Leahy. 2021.
\newblock \href {https://doi.org/10.48550/ARXIV.2101.00027} {The pile: An 800gb
  dataset of diverse text for language modeling}.

\bibitem[{Hey et~al.(2009)Hey, Tansley, Tolle, and
  Gray}]{hey2009-fourth-paradigm}
Tony Hey, Stewart Tansley, Kristin Tolle, and Jim Gray. 2009.
\newblock \href
  {https://www.microsoft.com/en-us/research/publication/fourth-paradigm-data-intensive-scientific-discovery/}
  {\emph{The Fourth Paradigm: Data-Intensive Scientific Discovery}}.
\newblock Microsoft Research.

\bibitem[{Hoffmann et~al.(2022)Hoffmann, Borgeaud, Mensch, Buchatskaya, Cai,
  Rutherford, de~Las~Casas, Hendricks, Welbl, Clark, Hennigan, Noland,
  Millican, van~den Driessche, Damoc, Guy, Osindero, Simonyan, Elsen, Rae,
  Vinyals, and Sifre}]{chinchilla-hoffmann-22}
Jordan Hoffmann, Sebastian Borgeaud, Arthur Mensch, Elena Buchatskaya, Trevor
  Cai, Eliza Rutherford, Diego de~Las~Casas, Lisa~Anne Hendricks, Johannes
  Welbl, Aidan Clark, Tom Hennigan, Eric Noland, Katie Millican, George van~den
  Driessche, Bogdan Damoc, Aurelia Guy, Simon Osindero, Karen Simonyan, Erich
  Elsen, Jack~W. Rae, Oriol Vinyals, and Laurent Sifre. 2022.
\newblock \href {https://doi.org/10.48550/arXiv.2203.15556} {Training
  compute-optimal large language models}.
\newblock \emph{CoRR}, abs/2203.15556.

\bibitem[{Hutchinson et~al.(2020)Hutchinson, Prabhakaran, Denton, Webster,
  Zhong, and Denuyl}]{hutchinson-etal-2020-social}
Ben Hutchinson, Vinodkumar Prabhakaran, Emily Denton, Kellie Webster, Yu~Zhong,
  and Stephen Denuyl. 2020.
\newblock \href {https://doi.org/10.18653/v1/2020.acl-main.487} {Social biases
  in {NLP} models as barriers for persons with disabilities}.
\newblock In \emph{Proceedings of the 58th Annual Meeting of the Association
  for Computational Linguistics}, pages 5491--5501, Online. Association for
  Computational Linguistics.

\bibitem[{Jo and Gebru(2020)}]{jo-archives-20}
Eun~Seo Jo and Timnit Gebru. 2020.
\newblock \href {https://doi.org/10.1145/3351095.3372829} {Lessons from
  archives: Strategies for collecting sociocultural data in machine learning}.
\newblock In \emph{Proceedings of the 2020 Conference on Fairness,
  Accountability, and Transparency}, FAT* '20, page 306–316, New York, NY,
  USA. Association for Computing Machinery.

\bibitem[{Kreutzer et~al.(2022)Kreutzer, Caswell, Wang, Wahab, van Esch,
  Ulzii-Orshikh, Tapo, Subramani, Sokolov, Sikasote, Setyawan, Sarin, Samb,
  Sagot, Rivera, Rios, Papadimitriou, Osei, Suarez, Orife, Ogueji, Rubungo,
  Nguyen, M{\"u}ller, M{\"u}ller, Muhammad, Muhammad, Mnyakeni, Mirzakhalov,
  Matangira, Leong, Lawson, Kudugunta, Jernite, Jenny, Firat, Dossou, Dlamini,
  de~Silva, {\c{C}}abuk~Ball{\i}, Biderman, Battisti, Baruwa, Bapna, Baljekar,
  Azime, Awokoya, Ataman, Ahia, Ahia, Agrawal, and
  Adeyemi}]{kreutzer-etal-2022-quality}
Julia Kreutzer, Isaac Caswell, Lisa Wang, Ahsan Wahab, Daan van Esch,
  Nasanbayar Ulzii-Orshikh, Allahsera Tapo, Nishant Subramani, Artem Sokolov,
  Claytone Sikasote, Monang Setyawan, Supheakmungkol Sarin, Sokhar Samb,
  Beno{\^\i}t Sagot, Clara Rivera, Annette Rios, Isabel Papadimitriou, Salomey
  Osei, Pedro~Ortiz Suarez, Iroro Orife, Kelechi Ogueji, Andre~Niyongabo
  Rubungo, Toan~Q. Nguyen, Mathias M{\"u}ller, Andr{\'e} M{\"u}ller,
  Shamsuddeen~Hassan Muhammad, Nanda Muhammad, Ayanda Mnyakeni, Jamshidbek
  Mirzakhalov, Tapiwanashe Matangira, Colin Leong, Nze Lawson, Sneha Kudugunta,
  Yacine Jernite, Mathias Jenny, Orhan Firat, Bonaventure F.~P. Dossou, Sakhile
  Dlamini, Nisansa de~Silva, Sakine {\c{C}}abuk~Ball{\i}, Stella Biderman,
  Alessia Battisti, Ahmed Baruwa, Ankur Bapna, Pallavi Baljekar, Israel~Abebe
  Azime, Ayodele Awokoya, Duygu Ataman, Orevaoghene Ahia, Oghenefego Ahia,
  Sweta Agrawal, and Mofetoluwa Adeyemi. 2022.
\newblock \href {https://doi.org/10.1162/tacl_a_00447} {Quality at a glance: An
  audit of web-crawled multilingual datasets}.
\newblock \emph{Transactions of the Association for Computational Linguistics},
  10:50--72.

\bibitem[{Krohs(2012)}]{krohs-convenience-experimentation}
Ulrich Krohs. 2012.
\newblock \href {https://doi.org/https://doi.org/10.1016/j.shpsc.2011.10.005}
  {Convenience experimentation}.
\newblock \emph{Studies in History and Philosophy of Science Part C: Studies in
  History and Philosophy of Biological and Biomedical Sciences}, 43(1):52--57.
\newblock Data-Driven Research in the Biological and Biomedical Sciences On
  Nature and Normativity: Normativity, Teleology, and Mechanism in Biological
  Explanation.

\bibitem[{Lauren{\c{c}}on et~al.(2022)Lauren{\c{c}}on, Saulnier, Wang, Akiki,
  del Moral, Scao, Werra, Mou, Ponferrada, Nguyen, Frohberg, {\v{S}}a{\v{s}}ko,
  Lhoest, McMillan-Major, Dupont, Biderman, Rogers, allal, Toni, Pistilli,
  Nguyen, Nikpoor, Masoud, Colombo, de~la Rosa, Villegas, Thrush, Longpre,
  Nagel, Weber, Mu{\~n}oz, Zhu, Strien, Alyafeai, Almubarak, Chien,
  Gonzalez-Dios, Soroa, Lo, Dey, Suarez, Gokaslan, Bose, Adelani, Phan, Tran,
  Yu, Pai, Chim, Lepercq, Ilic, Mitchell, Luccioni, and
  Jernite}]{laurencon2022the}
Hugo Lauren{\c{c}}on, Lucile Saulnier, Thomas Wang, Christopher Akiki,
  Albert~Villanova del Moral, Teven~Le Scao, Leandro~Von Werra, Chenghao Mou,
  Eduardo~Gonz{\'a}lez Ponferrada, Huu Nguyen, J{\"o}rg Frohberg, Mario
  {\v{S}}a{\v{s}}ko, Quentin Lhoest, Angelina McMillan-Major, G{\'e}rard
  Dupont, Stella Biderman, Anna Rogers, Loubna~Ben allal, Francesco~De Toni,
  Giada Pistilli, Olivier Nguyen, Somaieh Nikpoor, Maraim Masoud, Pierre
  Colombo, Javier de~la Rosa, Paulo Villegas, Tristan Thrush, Shayne Longpre,
  Sebastian Nagel, Leon Weber, Manuel~Romero Mu{\~n}oz, Jian Zhu, Daniel~Van
  Strien, Zaid Alyafeai, Khalid Almubarak, Vu~Minh Chien, Itziar Gonzalez-Dios,
  Aitor Soroa, Kyle Lo, Manan Dey, Pedro~Ortiz Suarez, Aaron Gokaslan, Shamik
  Bose, David~Ifeoluwa Adelani, Long Phan, Hieu Tran, Ian Yu, Suhas Pai, Jenny
  Chim, Violette Lepercq, Suzana Ilic, Margaret Mitchell, Sasha Luccioni, and
  Yacine Jernite. 2022.
\newblock \href {https://openreview.net/forum?id=UoEw6KigkUn} {The bigscience
  {ROOTS} corpus: A 1.6{TB} composite multilingual dataset}.
\newblock In \emph{Thirty-sixth Conference on Neural Information Processing
  Systems Datasets and Benchmarks Track}.

\bibitem[{Leonelli(2020)}]{leonelli-science-big-data}
Sabina Leonelli. 2020.
\newblock {Scientific Research and Big Data}.
\newblock In Edward~N. Zalta, editor, \emph{The {Stanford} Encyclopedia of
  Philosophy}, {S}ummer 2020 edition. Metaphysics Research Lab, Stanford
  University.

\bibitem[{Lhoest et~al.(2021)Lhoest, del Moral, Jernite, Thakur, von Platen,
  Patil, Chaumond, Drame, Plu, Tunstall, Davison, Sasko, Chhablani, Malik,
  Brandeis, Scao, Sanh, Xu, Patry, McMillan{-}Major, Schmid, Gugger, Delangue,
  Matussi{\`{e}}re, Debut, Bekman, Cistac, Goehringer, Mustar, Lagunas, Rush,
  and Wolf}]{huggingface-datasets-lhoest-21}
Quentin Lhoest, Albert~Villanova del Moral, Yacine Jernite, Abhishek Thakur,
  Patrick von Platen, Suraj Patil, Julien Chaumond, Mariama Drame, Julien Plu,
  Lewis Tunstall, Joe Davison, Mario Sasko, Gunjan Chhablani, Bhavitvya Malik,
  Simon Brandeis, Teven~Le Scao, Victor Sanh, Canwen Xu, Nicolas Patry,
  Angelina McMillan{-}Major, Philipp Schmid, Sylvain Gugger, Cl{\'{e}}ment
  Delangue, Th{\'{e}}o Matussi{\`{e}}re, Lysandre Debut, Stas Bekman, Pierric
  Cistac, Thibault Goehringer, Victor Mustar, Fran{\c{c}}ois Lagunas,
  Alexander~M. Rush, and Thomas Wolf. 2021.
\newblock \href {https://doi.org/10.18653/v1/2021.emnlp-demo.21} {Datasets: {A}
  community library for natural language processing}.
\newblock In \emph{Proceedings of the 2021 Conference on Empirical Methods in
  Natural Language Processing: System Demonstrations, {EMNLP} 2021, Online and
  Punta Cana, Dominican Republic, 7-11 November, 2021}, pages 175--184.
  Association for Computational Linguistics.

\bibitem[{Li et~al.(2023)Li, Allal, Zi, Muennighoff, Kocetkov, Mou, Marone,
  Akiki, Li, Chim, Liu, Zheltonozhskii, Zhuo, Wang, Dehaene, Davaadorj,
  Lamy{-}Poirier, Monteiro, Shliazhko, Gontier, Meade, Zebaze, Yee, Umapathi,
  Zhu, Lipkin, Oblokulov, Wang, V, Stillerman, Patel, Abulkhanov, Zocca, Dey,
  Zhang, Moustafa{-}Fahmy, Bhattacharyya, Yu, Singh, Luccioni, Villegas,
  Kunakov, Zhdanov, Romero, Lee, Timor, Ding, Schlesinger, Schoelkopf, Ebert,
  Dao, Mishra, Gu, Robinson, Anderson, Dolan{-}Gavitt, Contractor, Reddy,
  Fried, Bahdanau, Jernite, Ferrandis, Hughes, Wolf, Guha, von Werra, and
  de~Vries}]{starcoder}
Raymond Li, Loubna~Ben Allal, Yangtian Zi, Niklas Muennighoff, Denis Kocetkov,
  Chenghao Mou, Marc Marone, Christopher Akiki, Jia Li, Jenny Chim, Qian Liu,
  Evgenii Zheltonozhskii, Terry~Yue Zhuo, Thomas Wang, Olivier Dehaene, Mishig
  Davaadorj, Joel Lamy{-}Poirier, Jo{\~{a}}o Monteiro, Oleh Shliazhko, Nicolas
  Gontier, Nicholas Meade, Armel Zebaze, Ming{-}Ho Yee, Logesh~Kumar Umapathi,
  Jian Zhu, Benjamin Lipkin, Muhtasham Oblokulov, Zhiruo Wang, Rudra~Murthy V,
  Jason Stillerman, Siva~Sankalp Patel, Dmitry Abulkhanov, Marco Zocca, Manan
  Dey, Zhihan Zhang, Nour Moustafa{-}Fahmy, Urvashi Bhattacharyya, Wenhao Yu,
  Swayam Singh, Sasha Luccioni, Paulo Villegas, Maxim Kunakov, Fedor Zhdanov,
  Manuel Romero, Tony Lee, Nadav Timor, Jennifer Ding, Claire Schlesinger,
  Hailey Schoelkopf, Jan Ebert, Tri Dao, Mayank Mishra, Alex Gu, Jennifer
  Robinson, Carolyn~Jane Anderson, Brendan Dolan{-}Gavitt, Danish Contractor,
  Siva Reddy, Daniel Fried, Dzmitry Bahdanau, Yacine Jernite,
  Carlos~Mu{\~{n}}oz Ferrandis, Sean Hughes, Thomas Wolf, Arjun Guha, Leandro
  von Werra, and Harm de~Vries. 2023.
\newblock \href {https://doi.org/10.48550/arXiv.2305.06161} {Starcoder: may the
  source be with you!}
\newblock \emph{CoRR}, abs/2305.06161.

\bibitem[{Lin et~al.(2021)Lin, Ma, Lin, Yang, Pradeep, and
  Nogueira}]{Lin_etal_SIGIR2021_Pyserini}
Jimmy Lin, Xueguang Ma, Sheng-Chieh Lin, Jheng-Hong Yang, Ronak Pradeep, and
  Rodrigo Nogueira. 2021.
\newblock {Pyserini}: A {Python} toolkit for reproducible information retrieval
  research with sparse and dense representations.
\newblock In \emph{Proceedings of the 44th Annual International ACM SIGIR
  Conference on Research and Development in Information Retrieval (SIGIR
  2021)}, pages 2356--2362.

\bibitem[{McEnery and Hardie(2011)}]{mcenery_hardie_2011}
Tony McEnery and Andrew Hardie. 2011.
\newblock \href {https://doi.org/10.1017/CBO9780511981395} {\emph{Corpus
  Linguistics: Method, Theory and Practice}}.
\newblock Cambridge Textbooks in Linguistics. Cambridge University Press.

\bibitem[{Mitchell et~al.(2022)Mitchell, Luccioni, Lambert, Gerchick,
  McMillan{-}Major, Ozoani, Rajani, Thrush, Jernite, and
  Kiela}]{mitchell-measuring-data-22}
Margaret Mitchell, Alexandra~Sasha Luccioni, Nathan Lambert, Marissa Gerchick,
  Angelina McMillan{-}Major, Ezinwanne Ozoani, Nazneen Rajani, Tristan Thrush,
  Yacine Jernite, and Douwe Kiela. 2022.
\newblock \href {https://doi.org/10.48550/arXiv.2212.05129} {Measuring data}.
\newblock \emph{CoRR}, abs/2212.05129.

\bibitem[{MOI et~al.(2022)MOI, Patry, Cistac, Pete, Morgan, Pütz, Mishig,
  Johansen, Wolf, Gugger, Clement, Chaumond, Debut, Garillot, Georges, dctelus,
  Louis, MarcusGrass, Peyash, 0xflotus, deLevie, Mamaev, Arthur, Cameron,
  Clement, Moges, Hewitt, Zolotukhin, and Thomas}]{moi-hf-tokenizers}
Anthony MOI, Nicolas Patry, Pierric Cistac, Pete, Funtowicz Morgan, Sebastian
  Pütz, Mishig, Bjarte Johansen, Thomas Wolf, Sylvain Gugger, Clement, Julien
  Chaumond, Lysandre Debut, François Garillot, Luc Georges, dctelus, JC~Louis,
  MarcusGrass, Taufiquzzaman Peyash, 0xflotus, Alan deLevie, Alexander Mamaev,
  Arthur, Cameron, Colin Clement, Dagmawi Moges, David Hewitt, Denis
  Zolotukhin, and Geoffrey Thomas. 2022.
\newblock \href {https://doi.org/10.5281/zenodo.7298413}
  {huggingface/tokenizers: Rust 0.13.2}.

\bibitem[{Mökander et~al.(2023)Mökander, Schuett, Kirk, and
  Floridi}]{auditing-llms}
Jakob Mökander, Jonas Schuett, Hannah~Rose Kirk, and Luciano Floridi. 2023.
\newblock \href {https://doi.org/10.48550/ARXIV.2302.08500} {Auditing large
  language models: a three-layered approach}.

\bibitem[{Narayan et~al.(2018)Narayan, Cohen, and Lapata}]{xsum-emnlp}
Shashi Narayan, Shay~B. Cohen, and Mirella Lapata. 2018.
\newblock Don't give me the details, just the summary! {T}opic-aware
  convolutional neural networks for extreme summarization.
\newblock In \emph{Proceedings of the 2018 Conference on Empirical Methods in
  Natural Language Processing}, Brussels, Belgium.

\bibitem[{Nguyen et~al.(2016)Nguyen, Rosenberg, Song, Gao, Tiwary, Majumder,
  and Deng}]{msmarco}
Tri Nguyen, Mir Rosenberg, Xia Song, Jianfeng Gao, Saurabh Tiwary, Rangan
  Majumder, and Li~Deng. 2016.
\newblock \href {https://ceur-ws.org/Vol-1773/CoCoNIPS\_2016\_paper9.pdf} {{MS}
  {MARCO:} {A} human generated machine reading comprehension dataset}.
\newblock In \emph{Proceedings of the Workshop on Cognitive Computation:
  Integrating neural and symbolic approaches 2016 co-located with the 30th
  Annual Conference on Neural Information Processing Systems {(NIPS} 2016),
  Barcelona, Spain, December 9, 2016}, volume 1773 of \emph{{CEUR} Workshop
  Proceedings}. CEUR-WS.org.

\bibitem[{Niezni et~al.(2022)Niezni, Taub-Tabib, Harris, Sason-Bauer, Amrusi,
  Azagury, Avrashami, Launer-Wachs, Borchardt, Kusold, Tiktinsky, Hope,
  Goldberg, and Shamay}]{cancer-nlp-no-code}
Danna Niezni, Hillel Taub-Tabib, Yuval Harris, Hagit Sason-Bauer, Yakir Amrusi,
  Dana Azagury, Maytal Avrashami, Shaked Launer-Wachs, Jon Borchardt, M~Kusold,
  Aryeh Tiktinsky, Tom Hope, Yoav Goldberg, and Yosi Shamay. 2022.
\newblock \href {https://doi.org/10.1101/2022.05.03.490286} {Extending the
  boundaries of cancer therapeutic complexity with literature data mining}.
\newblock \emph{bioRxiv}.

\bibitem[{Ortiz~Su{\'a}rez et~al.(2019)Ortiz~Su{\'a}rez, Sagot, and
  Romary}]{suarez-processing-huge-corpora}
Pedro~Javier Ortiz~Su{\'a}rez, Beno{\^i}t Sagot, and Laurent Romary. 2019.
\newblock \href {https://doi.org/10.14618/IDS-PUB-9021} {{Asynchronous Pipeline
  for Processing Huge Corpora on Medium to Low Resource Infrastructures}}.
\newblock In \emph{{7th Workshop on the Challenges in the Management of Large
  Corpora (CMLC-7)}}, Cardiff, United Kingdom. {Leibniz-Institut f{\"u}r
  Deutsche Sprache}.

\bibitem[{pandas~development team(2020)}]{pandas-software}
The pandas~development team. 2020.
\newblock \href {https://doi.org/10.5281/zenodo.3509134} {pandas-dev/pandas:
  Pandas}.

\bibitem[{Piktus et~al.(2023)Piktus, Akiki, Villegas, Lauren{\c{c}}on, Dupont,
  Luccioni, Jernite, and Rogers}]{piktus-etal-2023-roots}
Aleksandra Piktus, Christopher Akiki, Paulo Villegas, Hugo Lauren{\c{c}}on,
  G{\'e}rard Dupont, Sasha Luccioni, Yacine Jernite, and Anna Rogers. 2023.
\newblock \href {https://aclanthology.org/2023.acl-demo.29} {The {ROOTS} search
  tool: Data transparency for {LLM}s}.
\newblock In \emph{Proceedings of the 61st Annual Meeting of the Association
  for Computational Linguistics (Volume 3: System Demonstrations)}, pages
  304--314, Toronto, Canada. Association for Computational Linguistics.

\bibitem[{Raffel et~al.(2022)Raffel, Shazeer, Roberts, Lee, Narang, Matena,
  Zhou, Li, and Liu}]{t5-raffel}
Colin Raffel, Noam Shazeer, Adam Roberts, Katherine Lee, Sharan Narang, Michael
  Matena, Yanqi Zhou, Wei Li, and Peter~J. Liu. 2022.
\newblock Exploring the limits of transfer learning with a unified text-to-text
  transformer.
\newblock \emph{J. Mach. Learn. Res.}, 21(1).

\bibitem[{Sherratt(2021)}]{glam-workbench}
Tim Sherratt. 2021.
\newblock \href {https://doi.org/10.5281/zenodo.5603060} {Glam workbench}.

\bibitem[{Siddiqui et~al.(2022)Siddiqui, Rajkumar, Maharaj, Krueger, and
  Hooker}]{hooker-archaelogy-22}
Shoaib~Ahmed Siddiqui, Nitarshan Rajkumar, Tegan Maharaj, David Krueger, and
  Sara Hooker. 2022.
\newblock \href {https://doi.org/10.48550/arXiv.2209.10015} {Metadata
  archaeology: Unearthing data subsets by leveraging training dynamics}.
\newblock \emph{CoRR}, abs/2209.10015.

\bibitem[{Vukovi{\'{c}} et~al.(2022)Vukovi{\'{c}}, Arora, Chang, Spitz, and
  West}]{Vukovi__2022}
Vuk Vukovi{\'{c}}, Akhil Arora, Huan-Cheng Chang, Andreas Spitz, and Robert
  West. 2022.
\newblock \href {https://doi.org/10.1145/3477495.3531696} {Quote erat
  demonstrandum: A web interface for exploring the quotebank corpus}.
\newblock In \emph{Proceedings of the 45th International {ACM} {SIGIR}
  Conference on Research and Development in Information Retrieval}. {ACM}.

\bibitem[{Weidinger et~al.(2021)Weidinger, Mellor, Rauh, Griffin, Uesato,
  Huang, Cheng, Glaese, Balle, Kasirzadeh, Kenton, Brown, Hawkins, Stepleton,
  Biles, Birhane, Haas, Rimell, Hendricks, Isaac, Legassick, Irving, and
  Gabriel}]{llm-risks-harms}
Laura Weidinger, John Mellor, Maribeth Rauh, Conor Griffin, Jonathan Uesato,
  Po{-}Sen Huang, Myra Cheng, Mia Glaese, Borja Balle, Atoosa Kasirzadeh, Zac
  Kenton, Sasha Brown, Will Hawkins, Tom Stepleton, Courtney Biles, Abeba
  Birhane, Julia Haas, Laura Rimell, Lisa~Anne Hendricks, William~S. Isaac,
  Sean Legassick, Geoffrey Irving, and Iason Gabriel. 2021.
\newblock \href {http://arxiv.org/abs/2112.04359} {Ethical and social risks of
  harm from language models}.
\newblock \emph{CoRR}, abs/2112.04359.

\bibitem[{{W}es {M}c{K}inney(2010)}]{pandas-paper}
{W}es {M}c{K}inney. 2010.
\newblock \href {https://doi.org/10.25080/Majora-92bf1922-00a} {{D}ata
  {S}tructures for {S}tatistical {C}omputing in {P}ython}.
\newblock In \emph{{P}roceedings of the 9th {P}ython in {S}cience
  {C}onference}, pages 56 -- 61.

\bibitem[{Zhang et~al.(2020)Zhang, Gupta, Tang, Han, Pradeep, Lu, Zhang,
  Nogueira, Cho, Fang, and Lin}]{zhang2020covidex}
Edwin Zhang, Nikhil Gupta, Raphael Tang, Xiao Han, Ronak Pradeep, Kuang Lu, Yue
  Zhang, Rodrigo Nogueira, Kyunghyun Cho, Hui Fang, and Jimmy Lin. 2020.
\newblock \href {https://doi.org/10.18653/v1/2020.sdp-1.5} {Covidex: Neural
  ranking models and keyword search infrastructure for the {COVID}-19 open
  research dataset}.
\newblock In \emph{Proceedings of the First Workshop on Scholarly Document
  Processing}, pages 31--41, Online. Association for Computational Linguistics.

\bibitem[{Zhang et~al.(2022)Zhang, Thakur, Ogundepo, Kamalloo, Alfonso-Hermelo,
  Li, Liu, Rezagholizadeh, and Lin}]{zhang-miracl-22}
Xinyu Zhang, Nandan Thakur, Odunayo Ogundepo, Ehsan Kamalloo, David
  Alfonso-Hermelo, Xiaoguang Li, Qun Liu, Mehdi Rezagholizadeh, and Jimmy Lin.
  2022.
\newblock \href {https://doi.org/10.48550/ARXIV.2210.09984} {Making a miracl:
  Multilingual information retrieval across a continuum of languages}.

\end{thebibliography}
\end{raggedright}

\end{document}